\documentstyle[12pt,epsf,epsfig,axodraw]{article}

\newcommand{\beq}{ \begin{eqnarray} }
\newcommand{\eeq}{ \end{eqnarray} }
\newcommand{\beqstar}{ \begin{eqnarray*} }
\newcommand{\eeqstar}{ \end{eqnarray*} }
\newcommand{\gsim}{ \mathop{}_{\textstyle \sim}^{\textstyle >} }
\newcommand{\lsim}{ \mathop{}_{\textstyle \sim}^{\textstyle <} }

\newcommand{\sla}[1]{\not\!#1}

\newcommand{\oneui}{\overline{\tilde{\chi}^0}}

\newcommand{\boxx}{{\rm box}}

\begin{document}
\baselineskip 0.7cm

\begin{titlepage}

\begin{center}

\hfill ICRR-Report-506-2004-4\\
\hfill YITP-04-39 \\
\hfill \today

{\large Direct Detection of the Wino- and Higgsino-like
Neutralino Dark Matters at One-Loop Level}
\vspace{1cm}

{\bf Junji Hisano}$^{1}$,
{\bf Shigeki Matsumoto}$^{1}$,
{\bf Mihoko M. Nojiri}$^{2}$
and 
{\bf Osamu Saito}$^{1}$
\vskip 0.15in
{\it
$^1${ICRR, University of Tokyo, Kashiwa 277-8582, Japan }\\
$^2${YITP, Kyoto University, Kyoto 606-8502, Japan}
}
\vskip 0.5in

\abstract{
The neutralino-nucleon ($\tilde{\chi}^0$-$N$) scattering is an
important process for direct dark matter searches.  In this paper we
discuss one-loop contributions to the cross section in the wino-like and
Higgsino-like LSP cases. The neutralino-nucleon scattering mediated by
the Higgs $\tilde{\chi}^0\tilde{\chi}^0$ and
$Z\tilde{\chi}^0\tilde{\chi}^0$ couplings at tree level is suppressed
by the gaugino-Higgsino mixing at tree level
when the neutralino is close to a weak eigenstate. The one-loop
contribution to the cross section, generated by the gauge interaction,
is not suppressed by any SUSY particle mass or mixing in the wino- and
Higgsino-like LSP cases.  It may significantly alter the total cross
section when $\sigma_{\tilde{\chi}^0 N}\sim 10^{-45}$ cm$^{2}$ or
less.  }

\end{center}
\end{titlepage}
\setcounter{footnote}{0}

\section{Introduction}

Dark matter mass density in the Universe is now measured very
precisely by cosmological observations, $\Omega_M=0.27\pm 0.04$
\cite{Spergel:2003cb}\cite{Bennett:2003bz}. Now one of the important 
questions regarding to the dark matter is the constituent. The minimal
supersymmetric standard model (MSSM) predicts the stable lightest
supersymmetric particle (LSP) if the R parity is conserved. This is
one of the attractive features of the model, because the neutralino
LSP is a good candidate of the dark matter in the Universe because 
it is a  weakly-interacting massive particle (WIMP).

The neutralino LSP is a linear combination of gauginos (bino and wino)
and Higgsinos, which are superpartners of the gauge and Higgs bosons in the SM,
respectively. The bino-like or Higgsino-like neutralino LSP is the dark
matter candidate in the minimal supergravity model (MSUGRA). Recently
many authors have investigated the cosmological relic density of the
bino-like dark matter in the MSUGRA. It is found that the thermal
relic density of the LSP is too large compared to the current
observations, unless the coannihilations with other  SUSY particles
enhance the effective neutralino annihilation cross section at the
early universe or entropy production reduces the number density after
the decoupling of the LSP.   The anomaly mediated SUSY-breaking model
\cite{Randall:1998uk,Giudice:1998xp} and  the string models with
moduli dominated SUSY breaking \cite{Brignole:1993dj} predict the
wino-like or Higgsino-like neutralino LSP. The thermal relic density of the
neutralino LSP is too low in the models unless the LSP mass
($m_{\tilde{\chi}^0}$) is above 1 TeV.  However, decay of
gravitino or other quasi-stable particles may produce the dark matter
non-thermally so that the relic abundance is consistent with the
observation \cite{Gherghetta:1999sw,Moroi:1999zb}. Also, while the LSP
with the mass heavier than about 1 TeV may lead to the naturalness
problem, it may be consistent in the split SUSY model
\cite{Arkani-Hamed:2004fb}.

Many experiments are now searching for the direct or indirect evidence
of the dark matter.  The counting rates in the direct search
experiments depend on the LSP neutralino-nucleon
($\tilde{\chi}^0$--$N$) interactions. The cross section above
$10^{-42}$ cm$^{2}$ is now explored for $m_{\tilde{\chi}^0}
\lsim 1$ TeV. While the annual modulation observed by the DAMA
experiment corresponds to the $\tilde{\chi}^0$--$N$ spin-independent
cross section around $10^{-42}$ cm$^{2}$
\cite{Bernabei:2003za}, the recent result of the CDMSII
rejects whole of the DAMA signal region \cite{Akerib:2004fq}
if the spin dependent part of the interaction is negligible.  The
sensitivity to the dark matter signal may be improved up to
$10^{-(45-46)}$ cm$^{2}$ or more in future.

The $\tilde{\chi}^0$--$N$ scattering cross section is sensitive to 
nature of the LSP and the SUSY particle mass spectrum. The Higgs and $Z$ boson
exchanges are dominant contributions to the spin-independent and
spin-dependent interactions responsible to the scattering,
respectively, in the wide parameter region. They are suppressed by the
gaugino-Higgsino mixing at tree level. When gauginos or Higgsino
is much heavier than the weak scale, the LSP is close to a pure weak
eigenstate and the scattering cross section is strongly
suppressed. The squark exchange also contributes to the cross section,
and it is not suppressed by the mixing. However, it tends to be
subdominant due to the heavier squark masses in the typical models.

In this paper, we evaluate the one-loop radiative corrections to the
wino- and Higgsino-like LSP scattering cross sections on a nucleon,
which are induced by the gauge interaction. These one-loop corrections
to the cross section are not suppressed by the Higgsino-gaugino
mixing. In addition to it, the loop integrals are only suppressed by
the weak gauge boson masses, because the chargino, which is the
$SU(2)$ partner of the LSP, is degenerated with the LSP in mass.  The
gauge-loop correction can dominate the total cross section in a limit
that the LSP is almost a pure weak eigenstate, setting the ``lower
limit'' of the total scattering cross section. When the LSP is
bino-like, the one-loop correction is negligible since bino does not
have gauge charges.

We note that similar phenomena sometimes appear in radiative
correction to the SUSY processes.  For example, it is known that the
mass difference between the LSP and chargino may be dominated by the
radiative correction due to the gauge loops in the case of the
wino-like LSP. The tree-level mass difference is
$O(m_Z^4/M_{SUSY}^3)$.  On the other hand, the radiative one is not
suppressed by any SUSY particle mass, and it is proportional to
$\alpha_2 m_W$. This is because the loop momentum around $m_W$
dominates in the loop integrals due to the mass degeneracy between the
LSP and the chargino.  In addition, the pair annihilation cross
sections of the wino- and Higgsino-like neutralino LSPs to two gammas at
one-loop level are not suppressed by the neutralino mass
\cite{Bergstrom:1997fh}. It is rather enhanced by a non-perturbative effect
when the mass is larger than the $m_W/\alpha_2$
\cite{Hisano:2003ec}. This effect also comes from the mass degeneracy between
the LSP and the chargino.

This paper is organized as follows. In Section 2, we set up the
formula for the general low-energy effective action and present the
$\tilde{\chi}^0$--$N$ scattering cross section. The interactions are
classified into the spin-dependent and spin-independent ones at the
non-relativistic limit of the neutralino LSP.  We note that there are
two classes for the spin-independent contributions; one is
proportional to the scalar operator of quark, $\langle N \vert
m_q\bar{q} q\vert N\rangle$, and the other is proportional to the
twist-2 operator, $\langle N\vert
\bar{q} i(\partial_{\mu} \gamma_{\nu}+
\partial_{\nu} \gamma_{\mu}-1/2 g_{\mu\nu} \sla{\partial})q 
\vert N \rangle $. 
The tree-level contribution to the twist-2 operators, which is induced
by the squark exchange, is negligible \cite{Drees:1993bu}. 
However, the twist-2 operators is sizable at one-loop level.

In Section 3, we briefly summarize the tree-level $\tilde{\chi}^0$
couplings responsible to the $\tilde{\chi}^0$--$N$ scattering for the
wino- and Higgsino-like LSPs.  In Section 4, the gauge-loop correction
to the effective action in the wino-like LSP case is presented. The
dominant contribution comes from the $W$ boson loops in this case.  We
identify the sources of the one-loop corrections to the
spin-independent interaction, correction to the Higgs coupling of the
LSP and those to scalar and twist-2 operators.  The
corrections are only suppressed by one-loop factors, not by the
Higgsino-gaugino mixing nor any SUSY particle mass. 
The one-loop correction in the Higgsino-like LSP case is summarized in
Appendix since the structure of the radiative correction 
is similar to the wino-like case.

In Section 5, we present some numerical results. Among the gauge-loop
contributions, sign of the correction to the twist-2 operator is
opposite to those to the scalar operator and Higgs boson vertex in the
case of the wino-like LSP. The cancellation reduces the total
correction to the spin-independent cross section in wide region of the
MSSM parameter space. Because of that, the spin-independent cross
section induced by the gauge loop alone is small, only around $\sim
10^{-(46-47)}$ cm$^2$ in the limit when the tree-level contribution to
the cross section is negligible. The total cross section, including
the contributions at tree and one-loop levels, will be affected by the
gauge-loop corrections when the cross section is close to the
sensitivities of the proposed dark matter search experiments
($\sigma_{\tilde{\chi}^0 N}\sim10^{-(45-46)}$cm$^2$). For the
Higgsino-like LSP, the cross section induced by the gauge-loop
diagrams alone is one order of magnitude smaller than that of the
wino-like LSP. Section 6 is devoted to conclusion and discussion.
    
\section{Effective Lagrangian for neutralino and nucleon scattering}

In this section we present the effective Lagrangian for
the $\tilde{\chi}^0$--$N$ scattering and the cross section formula.  
The effective interactions of the neutralino LSP with light quarks and
gluon at the renormalization scale $\bar{\mu}_0\simeq m_p$ are given in a
limit of the non-relativistic neutralino as follows
\begin{eqnarray}
{\cal L}^{\rm eff}&=&\sum_{q=u,d,s}{\cal L}^{\rm{eff}}_q +{\cal L}^{\rm{eff}}_g,
\end{eqnarray}
where
\beq 
{\cal L}^{\rm{eff}}_q
&=& 
d_q\ \oneui\gamma^{\mu}\gamma_5\tilde{\chi}^0\  \bar{q}\gamma_{\mu}\gamma_5 q
+
f_q m_q\ \oneui\tilde{\chi}^0\ \bar{q}q 
\cr
&+& \frac{g^{(1)}_q}{m_{\tilde{\chi}^0}} \ \oneui i \partial^{\mu}\gamma^{\nu} 
\tilde{\chi}^0 \ {\cal O}_{\mu\nu}^q
+ \frac{g^{(2)}_q}{m^2_{\tilde{\chi}^0}}\
\oneui(i \partial^{\mu})(i \partial^{\nu})
\tilde{\chi}^0 \ {\cal O}_{\mu\nu}^q
,
\label{eff_lagq}
\\
{\cal L}^{\rm eff}_{ g}&=&
f_G\ \oneui\tilde{\chi}^0 G_{\mu\nu}^aG^{a\mu\nu}
\nonumber\\
&+&\frac{g^{(1)}_G}{m_{\tilde{\chi}^0}}\
 \oneui i\partial^{\mu}\gamma^{\nu}
\tilde{\chi}^0 \ {\cal O}_{\mu\nu}^g
+
\frac{g^{(2)}_G}{m^2_{\tilde{\chi}^0}}\
\oneui(i\partial^{\mu}) (i\partial^{\nu})\tilde{\chi}^0 
\
{\cal O}_{\mu\nu}^g.
\label{eff_lagg}
\eeq
Here, we include terms up to the second derivative of the neutralino
field.  The first term of ${\cal L}^{\rm eff}_q$ is a spin-dependent
interaction, while the other terms in ${\cal L}^{\rm eff}_{q}$ and ${\cal
L}^{\rm eff}_{g}$ are spin-independent ``coherent'' contributions. The
third and fourth terms in ${\cal L}^{\rm eff}_q$ and the second and
third terms in ${\cal L}^{\rm eff}_g$ depend on the twist-2 operators
(traceless part of the energy momentum tensor) for 
quarks and gluon,
\beq
{\cal O}_{\mu\nu}^q&\equiv&\frac12 \bar{q} i \left(\partial_{\mu}\gamma_{\nu} +
\partial_{\nu}\gamma_{\mu}
-\frac{1}{2}g_{\mu\nu}\sla{\partial} \right)  q,
\nonumber\\
{\cal O}_{\mu\nu}^g&\equiv&\left(G_{~\mu}^{a~\rho}G_{~\rho\nu}^{a}+\frac{1}{4}g_{\mu\nu} 
G^a_{\alpha\beta}G^{a\alpha\beta}\right).
\eeq

The scattering cross section of the neutralino  with  target nuclei is
expressed compactly by using the coefficients given in ${\cal L}^{\rm
eff}_{q}$ and ${\cal L}^{\rm eff}_{g}$ as follows \cite{Jungman:1995df},
\begin{eqnarray}
\sigma&=&\frac{4}{\pi}\left(\frac{m_{\tilde{\chi}^0} m_T}{m_{\tilde{\chi}^0} +m_T}\right)^2
\left[(n_p f_p+n_nf_n)^2+4 \frac{J+1}{J} \left(
a_p\left\langle S_p\right\rangle+ a_n\left\langle S_n\right\rangle \right)^2\right].
\label{sigma}
\end{eqnarray}
The first
term in the bracket comes from the spin-independent interactions while
the second one is generated by the spin-dependent one.  

In Eq.~(\ref{sigma}), $m_T$ is the target nucleus mass and $n_p$ and
$n_n$ are proton and neutron numbers in the target nucleus,
respectively. The spin-independent coupling of the neutralino with
nucleon, $f_N~(N=p,n)$, in Eq.~(\ref{sigma}) is given as
\begin{eqnarray}
f_N/m_N&=&\sum_{q=u,d,s} \left(
f_qf_{Tq}+\frac{3}{4} (q(2,\bar{\mu}_0^2)+\bar{q}(2,\bar{\mu}_0^2))(g_q^{(1)}+g_q^{(2)})(\bar{\mu}_0) \right)
\nonumber\\
&-&\frac{8\pi}{9\alpha_s}f_{TG} f_G 
+\frac{3}{4} G(2,\bar{\mu}_0^2)\left(g^{(1)}_G
+g^{(2)}_G\right)(\bar{\mu}_0), \label{f}
\end{eqnarray}
where  the matrix elements of nucleon are expressed as 
\begin{eqnarray}
f_{Tq}&\equiv& \langle N \vert m_q \bar{q} q \vert N\rangle/m_N,
\nonumber
\\
f_{TG}&\equiv& 1-\sum_{u,d,s}f_{Tq},
\nonumber\\
\langle N(p)\vert 
{\cal O}_{\mu\nu}^q
\vert N(p) \rangle 
&=&\frac{1}{m_N}
(p_{\mu}p_{\nu}-\frac{1}{4}m^2_N g_{\mu\nu})\
(q(2,\bar{\mu}_0^2)+\bar{q}(2,\bar{\mu}_0^2)),
\nonumber\\
\langle N(p) \vert 
{\cal O}_{\mu\nu}^g
\vert N(p) \rangle
& =& \frac{1}{m_N}
(p_{\mu}p_{\nu}-\frac{1}{4}m^2_N g_{\mu\nu})\ 
G(2,\bar{\mu}_0).
\end{eqnarray}
Here,  $q(2,\bar{\mu}_0^2)$, $\bar{q}(2,\bar{\mu}_0^2)$  and  $G(2,\bar{\mu}_0^2)$ are the second moments of 
the quark, anti-quark and gluon distribution functions, respectively,
\begin{eqnarray}
q(2,\bar{\mu}^2_0)+ \bar{q}(2,\bar{\mu}^2_0) &=&\int^{1}_{0} dx ~x~ [q(x,\bar{\mu}^2_0)+\bar{q}(x,\bar{\mu}^2_0)],
\cr
G(2,\bar{\mu}^2_0) &=&\int^{1}_{0} dx ~x ~g(x,\bar{\mu}^2_0).
\end{eqnarray}

The constant $a_N$ ($N=p,n$), which is responsible for the
spin-dependent contribution, is defined as
\begin{eqnarray}
a_{N}&=&\sum_{q=u,d,s} d_q \Delta q_N,
\end{eqnarray}
\begin{eqnarray}
2 s_{\mu}\Delta q_N &\equiv& \langle N \vert 
\bar{q}\gamma_{\mu}\gamma_5  q \vert N \rangle,
\end{eqnarray}
where $s_{\mu}$ is the nucleon's spin, while $\langle S_N\rangle=
\langle A\vert S_N\vert A\rangle$ in Eq.~(\ref{sigma}) is 
the expectation value of 
the third component of the spin operator of the proton or neutron
group in the nucleus $A$.

The above formula does not contain heavy quark contributions  
explicitly.  After
integrating out  the Higgs and weak gauge bosons and squarks, 
the effective interactions of the neutralino with the heavy quarks
($Q$), which contribute to the spin-independent interaction, are
\begin{eqnarray}
{\cal L}^{\rm{eff}}_Q
&=& 
f_Q m_Q\ \oneui \tilde{\chi}^0\ \bar{Q}Q
+f_Q'\ \oneui \tilde{\chi}^0\ \bar{Q}i\sla{\partial} Q 
\cr
&+& \frac{g^{(1)}_Q}{m_{\tilde{\chi}^0}} \ \oneui i \partial^{\nu}\gamma^{\mu} 
\tilde{\chi}^0 \ {\cal O}_{\mu\nu}^Q
+ \frac{g^{(2)}_Q}{m^2_{\tilde{\chi}^0}}\
\oneui (i \partial^{\mu})(i \partial^{\nu})
\tilde{\chi}^0 \ {\cal O}_{\mu\nu}^Q.
\label{heavyquark}
\end{eqnarray}
The effective interactions (\ref{heavyquark}) contribute to ${\cal
L}^{\rm eff}_{g}$ through the heavy quark loop diagrams. 

It is well-known that the matrix element for $m_Q\bar{Q}Q$ in
Eq.~(\ref{heavyquark}) can be replaced by that for $-\alpha_s/(12\pi)
G^{a\mu\nu}G^a_{\mu\nu}$ due to the trace anomaly. Then,
\begin{eqnarray}
\langle N \vert m_Q \bar{Q} Q \vert N\rangle
&=&
\frac{2}{27}f_{TG}m_N.
\label{traceanomaly}
\end{eqnarray}
The second term of Eq.~(\ref{heavyquark}) is proportional to $\bar{Q}i\sla{\partial}Q$. It would reduce to $m_Q\bar{Q}Q
\oneui\tilde{\chi}^0$,
if the equation of motion for the heavy quark could be
applied. However, this is not justified because the heavy quark loop
diagram induced by the interaction, which contributes to the operator
$\oneui \tilde{\chi}^0 GG$, has an UV divergence.  Indeed it is found
in Ref.~\cite{Drees:1993bu}, where the squark exchange contribution to
the elastic scattering is evaluated, that an estimation of that
operator using the equation of motion for the heavy quarks disagrees
with the explicit full one-loop calculation by a factor of 2. We need
to calculate the vertex for $\oneui
\tilde{\chi}^0 GG$, which is induced by the heavy quark loop diagrams,
in the original theory. In this paper we parameterize the matrix
element of $\bar{Q} i\sla{\partial}Q$ as
\begin{eqnarray}
\langle N \vert \bar{Q}i\sla{\partial} Q \vert N\rangle
&=&
a_{\rm eff}~ \frac{2}{27}f_{TG}m_N.
\label{effectivepara}
\end{eqnarray}
by introducing a phenomenological parameter $a_{\rm eff}$. The precise
determination of $a_{\rm eff}$ requires evaluation of the higher-order
loop diagrams, and it is out of scope of this paper.

Eqs.~(\ref{eff_lagq}), (\ref{eff_lagg}) and (\ref{heavyquark}) contain
the traceless parts of the energy momentum tensor, $O_{\mu\nu}^q$,
$O_{\mu\nu}^Q$ and $O_{\mu\nu}^g$, whose matrix elements are
scale-dependent. The second moments of the quark and anti-quark
distribution functions, $q(2,\bar{\mu}^2)$ and $\bar{q}(2,\bar{\mu}^2)$, are mixed
with that of the gluon distribution function, $G(2,\bar{\mu}^2)$, once the
QCD radiative corrections are included. Their scale dependences are
compensated by those of the coefficients $g_q^{(1,2)}$ and
$g_G^{(1,2)}$, so that the total cross section is
scale-independent. Thus,
\begin{eqnarray}
\sum_{q=u,d,s} 
 (q(2,\bar{\mu}_0^2)+\bar{q}(2,\bar{\mu}_0^2))(g_q^{(1)}+g_q^{(2)})(\bar{\mu}_0) 
+ G(2,\bar{\mu}_0^2)  (g^{(1)}_G +g^{(2)}_G)(\bar{\mu}_0) 
\nonumber\\
=\sum_{m_q\le\bar{\mu}} 
(q(2,\bar{\mu}^2)+\bar{q}(2,\bar{\mu}^2))(g_q^{(1)}+g_q^{(2)})(\bar{\mu}) 
+ G(2,\bar{\mu}^2)(g^{(1)}_G 
+g^{(2)}_G)(\bar{\mu}). 
\end{eqnarray}
In this paper we use the second moments for gluon and quark
distribution functions at $\bar{\mu}=m_Z$, which are derived by the CTEQ
parton distribution
\cite{Pumplin:2002vw}.  

In Table~1, we show the parameters for the matrix elements, used in
this paper.  The second moments for the up and down quark distribution
functions are sizable.  We will see in Section 3, the term
proportional to $q(2,\bar{\mu}^2)$ is generated by the one-loop box
correction involving the $W$ boson exchanges.  The $f_{Tq}$ and
$f_{TG}$ represent the fractions of the trace part of the energy
momentum tensor, or the quark and gluino contributions to the nucleon
mass, as defined as above. The strange quark contribution to the
spin-independent cross section is dominant among those for light
quarks.  Also, the term proportional to $f_{TG}$ in Eq.~(\ref{f})
through heavy quark loops is not negligible.  

\begin{table}
\begin{center}
\begin{tabular}{|l|l|}
\hline
\multicolumn{2}{|c|}{For proton}\cr
\hline
$f_{Tu}$& 0.023\cr
$f_{Td}$& 0.034\cr
$f_{Ts}$&0.14\cr
\hline
\multicolumn{2}{|c|}{For neutron}\cr
\hline
$f_{Tu}$&0.019\cr
$f_{Td}$& 0.041\cr
$f_{Ts}$& 0.14 \cr
\hline
\end{tabular}
\hskip 1cm 
\begin{tabular}{|l|l|}
\hline
\multicolumn{2}{|c|}{Spin fraction}\cr
\hline
$\Delta u$& 0.77\cr
$\Delta d$& -0.49\cr
$\Delta s$& -0.15\cr
\hline 
\end{tabular}
\hskip 1cm 
\begin{tabular}{|l|l||l|l|}
\hline
\multicolumn{4}{|c|}{Second moment at $\bar{\mu}=m_Z$}\cr
\hline
$G(2)$&0.48&&\cr
$u(2)$&0.22&$\bar{u}(2)$& 0.034\cr
$d(2)$&0.11&$\bar{d}(2)$&0.036\cr
$s(2)$&0.026&$\bar{s}(2)$&0.026\cr
$c(2)$&0.019&$\bar{c}(2)$&0.019\cr
$b(2)$&0.012&$\bar{b}(2)$&0.012\cr
\hline
\end{tabular}
\end{center}
\caption{Parameters for quark and gluon matrix elements used in this paper. 
$f_{Ti}$ $(i=u,d,s)$ is taken from the estimation in
Refs.~\cite{Ashman:1989ig,Cheng:1988im,Jaffe:1989jz,Ellis:1991ef}.
The second moments for gluon and quarks at $\bar{\mu}=m_Z$ and the
spin fraction are for proton. Those for neutron are given by exchange
of up and down quarks in the tables. The second moments are calculated
using the CTEQ parton distribution
\cite{Pumplin:2002vw}. }
\label{table1}
\end{table}

\section{LSP elastic scattering induced by tree-level LSP couplings}

The mass matrix of the the neutralinos and
charginos are given by
\begin{eqnarray}
{\cal M}_N&=&\left(
\begin{array}{cccc}
M_1& 0 & -m_Zs_W \cos{\beta}& m_Z s_W \sin \beta\cr
0 & M_2 &m_Z c_W \cos{\beta}& -m_Z c_W \sin \beta\cr
-m_Z s_W \cos\beta& m_Zc_W \cos\beta& 0 &-\mu\cr
m_Z s_W \sin\beta& -m_Zc_W \sin\beta& -\mu & 0 \cr
\end{array}
\right)
\\
\cr
\cr
{\cal M}_C&=&
\left(\begin{array}{cc}
M_2 & \sqrt{2} m_W \sin\beta\cr
\sqrt{2}m_W \cos\beta & \mu\cr
\end{array}
\right),
\end{eqnarray}
which is written by the $(\tilde{B}, \tilde{W}^0, \tilde{H}^0_1, 
\tilde{H}^0_2)$ bases and $(\tilde{W}^+, \tilde{H}^+)$ bases, respectively.

When the lightest neutralino is wino-like ($M_2\ll\mu, M_1$) or
Higgsino-like ($\mu\ll M_1,M_2$), the mass eigenvalues of the lightest
neutralino and chargino are close to each other. They are given as
\begin{eqnarray}
m_{\tilde{\chi}^0}&=&M_2+\frac{m_W^2}{M_2^2-\mu^2}
(M_2 +\mu\sin2\beta)+...
\cr
m_{\tilde{\chi}^{-}}&=&M_2+\frac{m_W^2}{M_2^2-\mu^2}(M_2+\mu\sin 2\beta)
+...
\end{eqnarray}
for the wino case, and 
\begin{eqnarray}
m_{\tilde{\chi}^0}&=&
\mu+\frac{m_Z^2(1+\sin 2\beta)}{2(\mu-M_1)
(\mu-M_2)}(\mu-M_1 c_W^2-M_2 s_W^2)+...\cr
m_{\tilde{\chi}^{-}}&=&\mu-\frac{m_W^2}{M_2^2-\mu^2}(\mu+M_2\sin 2\beta)
+...
\end{eqnarray}
for the Higgsino case $(\mu>0)$. 
The lightest neutralino mass eigenstate becomes very close to the pure
wino or Higgsino in the limit, therefore $m_{\tilde{\chi}^0}\sim M_2$
(wino like) or $\sim \mu$ (Higgsino like).  The mass difference
between the LSP and the lighter chargino
$\delta_c=(m_{\tilde{\chi}^{-}}-m_{\tilde{\chi}^0})/
m_{\tilde{\chi}^0}$ is also very small if $||\mu|-M_2|\gg m_Z$. The
LSP and the lighter chargino form an $SU(2)$ triplet state in the
wino-like LSP case or vector-like doublets in the Higgsino-like one with
the second-lightest neutralino, respectively.

In this section, we discuss the tree-level contributions to the effective
interactions in Eqs.~(\ref{eff_lagq}), (\ref{eff_lagg}) and (\ref{heavyquark}) in the
wino-like and Higgsino-like LSP cases. The $\tilde{\chi}^0$--$N$
spin-independent interactions are generated by the $t$-channel
exchange of one Higgs boson and the $s$-channel exchange of one squark,
and the $\tilde{\chi}^0$--$N$ axial-vector interaction is induced by
the $t$-channel exchange of one $Z$ boson, respectively.  The
interactions of the neutralino LSP with the $Z$ and Higgs bosons are
suppressed by the mixing of gauginos and Higgsino. The squark exchange
also generates the twist-2 interaction. The twist-2 coupling is not
suppressed even in a pure gaugino limit, however, the term is
proportional to $m^{-4}_{\tilde{q}}$ in the amplitude. Overall, if the
other SUSY particles are much heavier than the LSP and the mixing of
gaugino and Higgsino is small, the elastic scattering is suppressed,
as mentioned in Introduction.  The one-loop contributions to the
$\tilde{\chi}^0$--$N$ scattering will be discussed in the next section.

\subsection{Spin-independent interaction}

The spin-independent interaction for the $\tilde{\chi}^0$--$N$
scattering arises from one Higgs boson or squark exchange at tree
level. Since the squark contribution is typically sub-dominant, we
concentrate on the Higgs boson contribution.

The neutralino coupling with a quark $q$, $f_q$, from the Higgs boson exchange is
given as
\begin{eqnarray}
f_q[H]&=&
\frac{g^2_2 }{4 m_W}
\left(
\frac{c_{h\tilde{\chi}\tilde{\chi}} c_{hqq}}{m^2_{h^0}}
+
\frac{c_{H\tilde{\chi}\tilde{\chi}} c_{Hqq}}{m^2_{H^0}}
\right),
\end{eqnarray}
where
\begin{eqnarray}
c_{hdd}=-\frac{\sin\alpha}{\cos\beta},&& 
c_{Hdd}= \frac{\cos\alpha}{\cos\beta}
\end{eqnarray}
for down-type quarks and 
\begin{eqnarray}
c_{huu}= \frac{\cos\alpha}{\sin\beta},&& 
c_{Huu}= \frac{\sin\alpha}{\sin\beta}
\end{eqnarray}
for up-type quarks. Here, $\alpha$ and $\beta$ are the mixing angle
of the neutral Higgs bosons and the vacuum mixing angle, 
and $m_{h^0}$ and $m_{H^0}$ are the light and heavy CP-even Higgs
boson masses, respectively. The tree-level coupling constants of the
neutralino LSP with the Higgs bosons, $c_{h\tilde{\chi}\tilde{\chi}}$
and $c_{H\tilde{\chi}\tilde{\chi}}$, are
\begin{eqnarray}
c_{h\tilde{\chi}\tilde{\chi}}
&=&
\left[(O_N)^\star_{12}-(O_N)^\star_{11}t_W\right]
\left[-\sin\alpha(O_N)^\star_{13}-\cos\alpha(O_N)^\star_{14}\right],\nonumber\\
c_{H\tilde{\chi}\tilde{\chi}}
&=&
\left[(O_N)^\star_{12}-(O_N)^\star_{11}t_W\right]
\left[\cos\alpha(O_N)^\star_{13}-\sin\alpha(O_N)^\star_{14}\right],
\label{hcc}
\end{eqnarray}
where $(O_N)$ is the neutralino mixing matrix. Here
$t_W=\tan\theta_W$, $c_W=\cos\theta_W$ and $s_W=\sin\theta_W$ with
$\theta_W$ the Weinberg angle.  

For the wino-like LSP, the couplings 
$c_{h\tilde{\chi}\tilde{\chi}}$ and $c_{H\tilde{\chi}\tilde{\chi}}$
are given as
\begin{eqnarray}
c_{h\tilde{\chi}\tilde{\chi}}\simeq
\frac{m_W}{M_2^2-\mu^2}(M_2+\mu\sin2\beta),
&&
c_{H\tilde{\chi}\tilde{\chi}}\simeq
-\frac{m_W}{M_2^2-\mu^2}\mu \cos2\beta.
\label{hccwino}
\end{eqnarray}
Here, we assume $||\mu|-M_2|\gg m_Z$ and $\cos\alpha\sim \sin\beta$
and $\sin\alpha\sim -\cos\beta$. The latter corresponds to a limit of
the heavy pseudoscalar Higgs boson mass ($m_A$).  The coupling
constants are suppressed when $|\mu|\gg M_Z$, as expected. When
$\tan\beta$ is large, the heavy Higgs boson contribution may be
dominant. While the LSP coupling with the light Higgs boson is
suppressed by $\sim m_{\tilde{\chi}^0} m_W/\mu^2$, the coupling of
strange quark to the heavy Higgs boson is enhanced proportional to
$\tan\beta$.

For the Higgsino-like LSP, we get
\begin{eqnarray}
c_{h\tilde{\chi}\tilde{\chi}}&\simeq&
\mp t_W^2 \frac12 \frac{m_W}{M_1-|\mu|}(1\pm \sin2\beta)
\mp \frac12 \frac{m_W}{M_2-|\mu|}(1\pm \sin2\beta),
\nonumber
\\
c_{H\tilde{\chi}\tilde{\chi}}&\simeq&
t_W^2 \frac12 \frac{m_W}{M_1-|\mu|} \cos 2\beta
+
\frac12 \frac{m_W}{M_2-|\mu|} \cos2\beta,
\label{Hcchiggs}
\end{eqnarray}
$\mu>0$ $(\mu<0)$. In this paper we take the LSP mass positive by an
axial rotation of the LSP field.  While the couplings are suppressed
by the gaugino masses, the suppression is moderate compared with the
wino-like LSP.

Assuming that the dominant contribution comes from the light Higgs
boson exchange, the cross section for 
the spin-independent $\tilde{\chi}^0$--$p$  scattering is 
approximately given as follows, 
\begin{eqnarray}
\sigma_{\rm SI}\sim
3\times 10^{-43}{\rm cm^2}\times 
\left(\frac{m_{h^0}}{115{\rm GeV}}\right)^{-4}
\left(\frac{\mu^2}{100{\rm GeV}\times M_2}\right)^{-2}
\left(1+\frac{\mu}{M_2} \sin2\beta\right)^2
\end{eqnarray}
for the wino-like LSP, and 
\begin{eqnarray}
\sigma_{\rm SI} \sim
1\times 10^{-43}{\rm cm^2}\times 
\left(\frac{m_{h^0}}{115{\rm GeV}}\right)^{-4}
\left(\frac{M_2}{100{\rm GeV}}\right)^{-2}
\end{eqnarray}
for the Higgsino-like LSP. Here we assume $M_2\ll\mu$ ($\mu \ll M_2=M_1)$ 
for the wino-like (Higgsino-like) neutralino LSP.
Note that the cross section for the the wino-like LSP
is suppressed by $\mu^{-4}$. For $\mu=1$ TeV and $M_2=100$ GeV and
$\tan\beta=10$, the cross section reduces down to $10^{-46}$ to
$10^{-47}$ cm$^{2}$.  It is found in Ref.~\cite{profumo}  that
the wino- and Higgsino-like LSP masses are about (1-2) TeV and the
spin-independent cross sections are $10^{-(44-48)}$cm$^2$, when
imposing the thermal relic density constraint.

\subsection{Spin-dependent interaction}

The spin-dependent interaction of the LSP arises from one $Z$ boson or
one squark exchange. We ignore the squark contribution here, again.  The
tree-level contribution to $d_q$ in Eq.~(\ref{eff_lagq}) from the  $Z$
boson exchange is represented as 
\begin{eqnarray}
d_q&=&
\frac{g^2_2}{8m_W^2}T_3^q c_{Z\tilde{\chi}\tilde{\chi}},
\end{eqnarray}
where $T_3^q$ is for the isospin of a quark $q$. The tree-level
LSP coupling to $Z$ boson, $c_{Z\tilde{\chi}\tilde{\chi}}$, is
\begin{eqnarray}
c_{Z\tilde{\chi}\tilde{\chi}}
&=&
\vert (O_N)_{13}\vert^2-\vert (O_N)_{14}\vert^2.
\end{eqnarray}

For the wino-like  LSP, $c_{Z\tilde{\chi}\tilde{\chi}}$ becomes
\begin{eqnarray}
c_{Z\tilde{\chi}\tilde{\chi}}
&\simeq&
\frac{m_W^2}{M_2^2-\mu^2} \cos2\beta.
\label{wino_z}
\end{eqnarray}
On the other hand, in a limit of the Higgsino-like LSP,
\begin{eqnarray}
c_{Z\tilde{\chi}\tilde{\chi}}
&\simeq&
\mp \frac12 \left(t_W^2 \frac{m_W^2}{M_1\mu}+\frac{m_W^2}{M_2\mu}\right)
\cos2\beta +O\left(\frac{\mu}{M_1},\frac{\mu}{M_2}\right),
\label{higg_z}
\end{eqnarray}
for $\mu>0$ $(\mu<0)$. Thus, again, the coupling for the wino-like LSP is
more suppressed than that for the Higgsino-like one. 

Using Eqs.~(\ref{wino_z}) and (\ref{higg_z}), the spin-dependent 
cross section with proton is approximately given as 
\begin{eqnarray}
\sigma_{\rm SD}\sim
2\times 10^{-38}{\rm cm^2}\times 
\left(\frac{\mu}{100{\rm GeV}}\right)^{-4}
\cos^22\beta
\label{sigmasdtreew}
\end{eqnarray}
for the wino-like LSP, and 
\begin{eqnarray}
\sigma_{\rm SD}\sim
8\times 10^{-39}{\rm cm^2}\times 
\left(\frac{M_2}{100{\rm GeV}}\right)^{-2}
\left(\frac{\mu}{100{\rm GeV}}\right)^{-2}
\cos^22\beta
\label{sigmasdtreeh}
\end{eqnarray}
for the Higgsino-like LSP.  Here, we take $M_1=M_2$.
The analysis  in Ref.~\cite{profumo} shows that 
the spin-dependent
cross sections are $10^{-(41-45)}$cm$^2$ for the wino- and Higgsino-like LSPs
when imposing the thermal relic density constraint.

\section{Elastic scattering induced by one-loop effective action }

In the previous section, we discuss that the interactions responsible
for the $\tilde{\chi}^0$--$N$ scattering are suppressed by the
gaugino-Higgsino mixing or squark masses at tree level. However, this
is not true for the radiative corrections to the effective
interactions if the dark matter is wino- or Higgsino-like, because of
the mass degeneracy between the LSP and its $SU(2)$ partner. In this
section, we derive radiative corrections to the effective
interactions in Eqs.~(\ref{eff_lagq}), (\ref{eff_lagg}) and
(\ref{heavyquark}), and it is found that some of them are only
suppressed by the weak gauge boson mass at most.

We first discuss the anomalous Higgs boson  vertices of the neutralino and
the box diagram contributions involving the $W$ bosons to the effective
interactions for the case of the wino-like LSP. The numerical result
will be shown in the next section.  For the case of the Higgsino-like
LSP, we present the explicit formula for the radiative corrections in
Appendix.

\begin{figure}
\begin{center}
\begin{picture}(455,140)(30,-20)
\PhotonArc(135,25)(65,0,180){2}{20}
\large
\Text(135,105)[]{$W^-$}

\Line(60,25)(30,25)
\Line(90,25)(60,25)
\Line(135,25)(90,25)
\Line(135,25)(180,25)
\Line(180,25)(210,25)
\Line(210,25)(240,25)

\Text(45,15)[]{\large$\tilde{\chi}^0$}
\Text(135,15)[]{$\tilde{\chi}^-$}
\Text(225,15)[]{$\tilde{\chi}^0$}

\DashLine(172,100)(190,130){3}
\Text(203,145)[]{$h^0$, $H^0$}

\Text(135,-10)[]{(a)}

\SetOffset(245,0)

\Line(60,25)(45,25)
\Line(110,25)(60,25)
\Line(110,25)(160,25)
\Line(160,25)(210,25)
\Line(210,25)(225,25)

\ArrowLine(45,125)(60,125)
\ArrowLine(60,125)(210,125)
\ArrowLine(210,125)(225,125)

\Photon(90,125)(90,25){2}{10}
\Photon(180,125)(180,25){2}{10}

\Text(60,15)[]{$\tilde{\chi}^0$}
\Text(135,15)[]{$\tilde{\chi}^-$}
\Text(210,15)[]{$\tilde{\chi}^0$}

\Text(70,75)[]{${W}^-$}
\Text(205,75)[]{${W}^-$}

\Text(60,110)[]{$q$}
\Text(135,110)[]{$q'$}
\Text(210,110)[]{$q$}

\Text(135,-10)[]{(b)}
d
\end{picture} 

\end{center}
\caption{
(a) Diagram contributing to the anomalous Higgs  boson vertices of the
neutralino and (b) box diagram contributing to the
$\tilde{\chi}^0$--$N$ scattering in the case of the wino-like LSP.}
\label{fig1}
\end{figure}
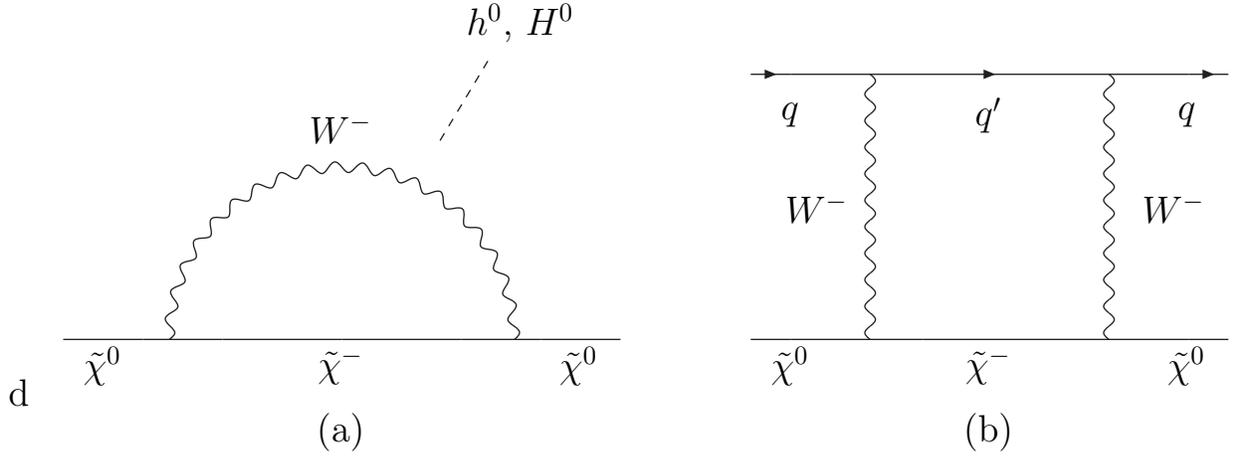

The gauge interactions of the wino-like neutralino and chargino are
\begin{eqnarray}
  {\cal L}_{\rm int}
  &=&
  -\frac{e}{s_W}
  \left(
    \overline{\tilde{\chi}^0}\gamma^\mu\tilde{\chi}^-W^\dagger_\mu
    +
    h.c.
  \right)
  +
  e \frac{c_W}{s_W}\overline{\tilde{\chi}^-}\gamma^\mu\tilde{\chi}^-Z_\mu
  +
  e\overline{\tilde{\chi}^-}\gamma^\mu\tilde{\chi}^-A_\mu~.
\end{eqnarray}
Here we ignore the mixings of the neutralinos and charginos for
simplicity.  These interactions induce the anomalous Higgs boson
vertices and the box diagrams which contribute to the
$\tilde{\chi}^0$--$N$ scattering. The Feynman diagrams are given in
Fig.~1.  The radiative correction to the $Z$ boson vertex is not
induced at the one-loop level due to the Majorana nature of the LSP.

We start from the radiative correction to the the Higgs boson
vertices. The tree-level contribution is given in Eq.~(\ref{hcc}).
The radiative corrections to
$c_{h\tilde{\chi}\tilde{\chi}}$ and $c_{H\tilde{\chi}\tilde{\chi}}$ at
 one-loop level are expressed as
\begin{eqnarray}
\delta c_{h\tilde{\chi}\tilde{\chi}}
&=&
\frac{\alpha_2}{4\pi}
\sin({\alpha-\beta})
\left[F_{\rm H}^{(0)}(x_W)
+
\delta_C
F_{\rm H}^{(1)}(x_W)
\right],
\\
\delta c_{H\tilde{\chi}\tilde{\chi}}
&=&
-\frac{\alpha_2}{4\pi}
\cos({\alpha-\beta})
\left[F_{\rm H}^{(0)}(x_W)
+
\delta_C
F_{\rm H}^{(1)}(x_W)
\right],
\end{eqnarray}
respectively.
The correction to $f_q[H](= f^{tree}_q[H]+\delta f_q[H])$
is therefore 
\begin{equation} 
\delta f_q[H]=\frac{g^2}{4m_W} \left(
\frac{\delta c_{h\tilde{\chi}\tilde{\chi}} c_{hqq}}{m_{h^0}^2}
+\frac{\delta c_{H\tilde{\chi}\tilde{\chi}} c_{Hqq}}{m_{H^0}^2}
\right).
\label{higgs_c}
\end{equation}
Here, $x_W=m_W^2/m^2_{\tilde{\chi}^0}$ and
$\bar{b}_W=\sqrt{1-x_W/4}$. We expand the radiative corrections by
$\delta_C\equiv (m_{\tilde{\chi}^-}
-m_{\tilde{\chi}^0})/m_{\tilde{\chi}^0}$.  When the gaugino and
Higgsino masses are comparable to the weak gauge boson masses, this
expansion of $\delta_C$ is invalid, however, the one-loop contributions
to the cross section are negligible. We are interested in a case
where  the gaugino or
Higgsino mass is much larger than the weak gauge boson masses
and the LSP is close to the Higgsino or wino weak eigenstate.
This expansion is justified in the case.

The mass functions are
\begin{eqnarray}
F_{\rm H}^{(0)}(x)
&=&
\frac{2}{\bar{b}_W}
(2+x(2-x))\tan^{-1}(\frac{2\bar{b}_W}{\sqrt{x}})
-
2\sqrt{x}(2-x \log(x)),
\\
F_{\rm H}^{(1)}(x)
&=&
\frac{6}{\bar{b}_W^3}
\tan^{-1}(\frac{2\bar{b}_W}{\sqrt{x}})
-
\frac{2}{\bar{b}_W^2}\frac{1}{\sqrt{x}}(2+x).
\end{eqnarray}
Since $F_{\rm H}^{(0)}(x)$ becomes $2\pi$ in a limit of $x\rightarrow 0$,
the correction does not vanish in the heavy neutralino limit. When the
pseudoscalar Higgs boson mass is large, $\delta
c_{h\tilde{\chi}\tilde{\chi}}$ is maximum and $\delta
c_{H\tilde{\chi}\tilde{\chi}}$ is vanishing.

The effective interactions in  Eqs.~(\ref{eff_lagq}) and (\ref{heavyquark})
receive the radiative correction from the the box diagrams (Fig.~1(b)) at
the renormalization scale $\bar{\mu}\simeq m_W$ of the form,
\begin{eqnarray}
\delta {\cal L}^{\rm{eff}}_q[\boxx]
&=& \delta d_q[\boxx]\ 
\overline{\tilde{\chi}^0}\gamma^{\mu}\gamma_5\tilde{\chi}^0\  \bar{q}\gamma_{\mu}\gamma_5 q
\nonumber\\
&+&
\delta f_q[\boxx]\  m_q\overline{\tilde{\chi}^0}\tilde{\chi}^0\  \bar{q}q 
+\delta f'_q[\boxx]\  \overline{\tilde{\chi}^0}\tilde{\chi}^0 \  \bar{q}i\sla{\partial} q 
\cr
&+& \frac{\delta g^{(1)}_q[\boxx]}{m_{\tilde{\chi}^0}} \ \overline{\tilde{\chi}^0} i \partial^{\mu}\gamma^{\nu} 
\tilde{\chi}^0 \ {\cal O}_{\mu\nu}^q
+ \frac{\delta g^{(2)}_q[\boxx]}{m^2_{\tilde{\chi}^0}}\overline{\tilde{\chi}^0}
(i \partial^{\mu})(i \partial^{\nu})
\tilde{\chi}^0 \ {\cal O}_{\mu\nu}^q,
\label{boxeff}
\end{eqnarray}
where  
\begin{equation}
\delta d_q[\boxx]=\frac{\alpha_2^2}{m_W^2} F_{\rm AV}(x_W),\
\delta f_q[\boxx]=\frac{\alpha_2^2}{m_W^3}F_{S1}(x_W), \
\delta f'_q[\boxx]=\frac{\alpha_2^2}{m_W^3}F_{S2}(x_W), 
\end{equation}
\begin{equation}
\delta g_q^{(1)}[\boxx]=\frac{\alpha_2^2}{m_W^3} F_{T1}(x_W),
\ 
\delta g_q^{(2)}[\boxx]=\frac{\alpha_2^2}{m_W^3} F_{T2}(x_W).
\end{equation}
In our calculation of the radiative corrections, we ignore $O(m_q^2)$,
and expand the loop integrals up to order of $p$ and $\delta_C$, where
$p$ is the quark external momentum. In this approximation the loop
integrals are expressed analytically by $B$ functions and its
derivatives \cite{'tHooft:1978xw}. This procedure is not justified for
box diagrams with the external or internal top quarks since it is
heavier than the weak gauge bosons, however, the radiative corrections
should be suppressed by the top quark mass. Thus, they are sub-dominant
compared with the other lighter quark ones.

The loop function can be expanded as $F_I(x)= F_I^{(0)}(x)+\delta_C
F_I^{(1)}(x)$ $(I=$  AV, S1, S2, T1, and T2), and they are
given as follows;
\begin{eqnarray}
F^{(0)}_{\rm AV}(x)
&=&
\frac{1}{24\bar{b}_W}\sqrt{x}(8-x-x^2)\tan^{-1}(\frac{2\bar{b}_W}{\sqrt{x}})
-\frac{1}{24} x(2-(3+x)\log(x)),
\nonumber\\
F^{(1)}_{\rm AV}(x)
&=&
\frac{1}{4\bar{b}_W^3}\sqrt{x}\tan^{-1}(\frac{2\bar{b}_W}{\sqrt{x}})
-\frac{1}{2\bar{b}_W^2},
\nonumber\\
F^{(0)}_{\rm S1}(x)
&=&
F^{(1)}_{\rm S1}(x)
=0,
\nonumber\\
F^{(0)}_{\rm S2}(x)
&=&
-\frac{\bar{b}_W}{24}(2+x^2) \tan^{-1}(\frac{2\bar{b}_W}{\sqrt{x}})
-\frac{1}{96} \sqrt{x}(1-2x-x(2-x) \log(x)),
\nonumber\\
F^{(1)}_{\rm S2}(x)
&=&
-\frac{1}{24\bar{b}_W}(1-x)^2
 \tan^{-1}(\frac{2\bar{b}_W}{\sqrt{x}})
\nonumber\\
&&
+\frac{1}{24\sqrt{x}}(
2(6-x)+6\log(2\delta_C)-(3-x^2) \log(x)),
\nonumber\\
F^{(0)}_{\rm T1}(x)
&=&
\frac{1}{6}\bar{b}_W(2+x^2)\tan^{-1}(\frac{2\bar{b}_W}{\sqrt{x}})
+\frac{1}{24}\sqrt{x}(1-2x-x(2-x)\log{x}), 
\nonumber\\
F^{(1)}_{\rm T1}(x)
&=&
\frac{1}{6\bar{b}_W}(1-2x+x^2)\tan^{-1}(\frac{2\bar{b}_W}{\sqrt{x}})
\nonumber\\
&&
+\frac{1}{6\sqrt{x}}
(2x+6\log(2\delta_C)-(3+x^2)\log(x)),
\nonumber\\
F^{(0)}_{\rm T2}(x)
&=&
\frac{1}{4\bar{b}_W} x (2-4x+x^2)\tan^{-1}(\frac{2\bar{b}_W}{\sqrt{x}})
-\frac{1}{4}\sqrt{x}(1-2x-x(2-x)\log(x)),
\nonumber\\
F^{(1)}_{\rm T2}(x) &=&
\frac{1}{24\bar{b}_W^3}(16+30x-30x^2+5x^3)
\tan^{-1}(\frac{2\bar{b}_W}{\sqrt{x}})
\nonumber\\
&&
-\frac{1}{24\bar{b}_W^2}
\sqrt{x}(28-10x-5x(4-x)\log(x)).
\label{massfunc}
\end{eqnarray}
Note that $F_{\rm S1}(x)$ is zero due to the chiral nature of $W\bar{q}q$
vertex. For the Higgsino-like LSP, it does not vanish because 
the $Z$ boson  couples with  both $q_L$ and $q_R$.  See Appendix. 

The functions $F^{(0)}_{\rm S2}(x)$ and $F^{(0)}_{\rm T1}(x)$ are
non-vanishing even if $x$ approaches to 0 (or $m_{\tilde{\chi}^0}$ is
increased), and they become $-\pi/24$ and $\pi/6$,
respectively. Thus, in addition to the correction to the Higgs boson
vertices in Eq.~(\ref{higgs_c}), those to the scalar and twist 2
operators induced by the $W$ boson loops do not vanish in a heavy
LSP limit, and they contribute to the spin-independent cross
section. On the other hand, the spin-dependent cross section depends
on $F_{\rm AV}(x)$, which is suppressed in the heavy LSP limit as
$\propto m_W/m_{\tilde{\chi}^0}$.

The functions, $F^{(1)}_{\rm H}(x)$, $F^{(1)}_{\rm S2}(x)$ and $F^{(1)}_{\rm
T1}(x)$, are proportional to $1/\sqrt{x}$ for small $x$. This does
not cause a problem. The mass difference $\delta_C$ is proportional to
$x_W^{3/2}$ at tree level, and it becomes proportional to $\alpha_2
x_W^{1/2}$ due to the radiative correction when the LSP mass is much
larger than the $W$ boson mass. Thus, $\delta_C/\sqrt{x_W} \sim x_{W}$ or
$\sim \alpha_2$.  Therefore the perturbation by $\delta_C$ is not
broken.

\begin{figure}
\begin{center}
\begin{picture}(255,170)(30,-20)
\large
\Line(80,25)(30,25)
\Photon(145,25)(80,25){2}{10}
\Photon(145,125)(80,125){2}{10}
\Line(145,25)(145,125)

\Line(145,125)(210,125)
\Line(145,25)(210,25)
\Line(210,125)(210,25)

\Line(145,25)(210,25)
\Gluon(210,25)(260,25){3}{5}
\Gluon(210,125)(260,125){3}{5}
\ArrowLine(30,125)(60,125)
\Line(60,125)(80,125)
\Line(80,125)(80,25)

\Text(45,15)[]{$\tilde{\chi}^0$}
\Text(120,15)[]{$W^-$}
\Text(235,15)[]{$g$}

\Text(65,75)[]{$\tilde{\chi}^-$}
\Text(220,75)[]{$q$}
\Text(155,75)[]{$q'$}
\Text(185,110)[]{$q$}
\Text(185,15)[]{$q$}

\Text(45,110)[]{$\tilde{\chi}^0$}
\Text(120,110)[]{$W^-$}
\Text(235,110)[]{$g$}


\large

\end{picture}
\end{center}
\caption{A two-loop diagram contributing to $GG\overline{\tilde{\chi}^0}\tilde{\chi}^0$.}
\label{2loop}
\end{figure}
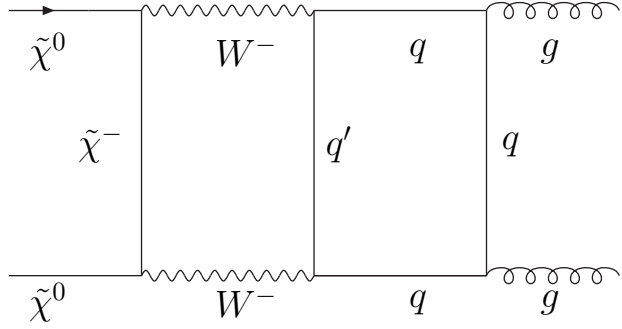

\section{Numerical results}

In the previous sections, we derived the formulas for the one-loop
corrections to the effective interactions. We now calculate the cross
section following the procedure in Section 2.

First, we discuss the cross section for the spin-independent
$\tilde{\chi}^0$--$p$ scattering induced by the gauge-loop diagrams
presented in the previous section, assuming the tree-level
contribution to the cross section is negligible. The total cross
section involving both the tree and one-loop level contributions  is
more sensitive to the MSSM model parameters, and it is discussed
later.

For the wino-like LSP, the spin-independent interaction induced by
the $W$ boson at one-loop level is approximately represented by
\begin{eqnarray}
\delta f_N
&=&
-\frac{\pi \alpha_2^2 m_N}{m_W m_{h^0}^2}\left(\sum_{q=u,d,s}\frac{f_{Tq}}{2}+\frac{f_{TG}}{9}\right)
-\frac{\pi \alpha_2^2 m_N}{m_W^3}\left(\sum_{q=u,d,s}\frac{f_{Tq}}{24}+a_{\rm eff} \frac{f_{TG}}{324}\right)
\nonumber\\
&&+\sum_{q=u,d,s,c}\frac{\pi \alpha_2^2 m_N}{8 m_W^3} (q(2,m_W^2)+\bar{q}(2,m_W^2)).
\label{fn}
\end{eqnarray}
Here, we take the pseudoscalar Higgs boson and the LSP to be much heavier
than the $W$ boson mass. The first term comes from the correction to
the light Higgs boson vertex, proportional to $\delta
c_{h\tilde{\chi}\tilde{\chi}}$. The second one is from those to the
scalar operators and it is given by $\delta f_q$ and $\delta
{f^{'}}_q$ in Eq. (\ref{boxeff}). The third (last) one represents
those to the twist-2 operators, which are proportional to $\delta
g^{(1)}_q$ and $\delta g^{(2)}_q$ in Eq.~(\ref{boxeff}). Here, we
include the top and bottom quark contributions to the Higgs vertex
correction, while those to the scalar and twist-2 operators, which are
suppressed by the top quark mass, are ignored. As discussed before,
$a_{\rm eff}$ is a phenomenological parameter for $\langle N \vert
\bar{Q} i\sla{\partial} Q\vert N\rangle$. Eq.~(\ref{fn}) is
numerically given as 
(in the order of the terms in Eq.~(\ref{fn})) 
\begin{eqnarray}
\delta f_N
&=& -5.8\times 10^{-10} -(5.3+1.6 a_{\rm eff})\times 10^{-11}
+3.9\times 10^{-10} ({\rm GeV^{-2}})
\nonumber\\
&=&-2.6\times 10^{-10} ({\rm GeV^{-2}}), ~~~(a_{\rm eff}=1).
\label{order}
\end{eqnarray}
Here, $m_{h^0}=115$~GeV and the parameters in Table~\ref{table1} for
the hadronic matrix elements are used. We assume $a_{\rm eff}=1$ in
the second line. We found the contributions from the terms with
different spin structure cancel each other. This value for $\delta
f_N$ corresponds to $\sigma_{\rm SI}\simeq 3.0\times 10^{-47}{\rm
cm}^2$, if the tree-level contribution is negligible. The correction
is still large enough to alter the total cross section close to the
proposed sensitivities for the future experiments ($10^{-(45-46)}$
cm$^2$). Here, we present the spin-independent interaction of the
neutralino with proton, however, the value for that with neutron is
almost the same as it since the numerical difference comes only from
the Higgs exchange contributions.

\begin{figure}
 \centerline{\epsfxsize = 0.7\textwidth \epsffile{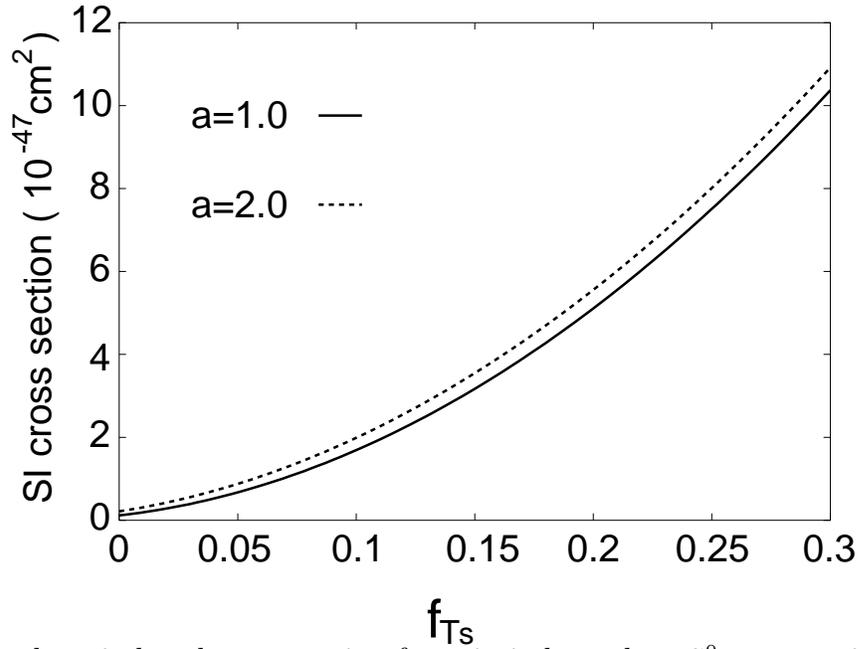} }
\vspace*{-5mm}
\caption{One-loop induced cross
section for spin-independent $\tilde{\chi}^0$--$p$ scattering as
a function of $f_{Ts}$ for $a_{\rm eff}=1.0$ and 2.0 in the case of the
wino-like LSP. We assume that the tree-level contribution is
negligible and the process is induced by the one-loop diagrams.  Here,
we take $m_{\tilde{\chi}^0}=1600$GeV and the parameter in
Table.~\ref{table1} except for $f_{Ts}$.  Also, we use $\tan\beta=10$,
$m_A$=1000GeV, $m_{\rm stop}$=2000GeV, and $A_{\rm top}$=0, which
determine the light Higgs mass and the mixing angle.  }
\label{outputa}
\end{figure}

In Fig.~\ref{outputa}, we show the one-loop induced cross section for
the spin-independent $\tilde{\chi}^0$--$p$ scattering as a function of
$f_{Ts}$ for $a_{\rm eff}=1.0$ and 2.0. The
tree-level contribution is assumed to be negligible. The parameters
$f_{Ts}$ and $a_{\rm eff}$ have large theoretical uncertainties. The
precise determination of $a_{\rm eff}$ requires calculation of 
two-loop diagrams, such as in Fig.~\ref{2loop}.  Here, we take
$m_{\tilde{\chi}^0}=1600$GeV and the parameters in Table.~\ref{table1}
except for $f_{Ts}$.  Also, we use  FeynHiggs
\cite{Heinemeyer:1998yj} in order to
calculate the Higgs boson masses and mixing angle.  In this figure, we
use $\tan\beta=10$, $m_A$=1000GeV, $m_{\rm stop}$=2000GeV, and $A_{\rm
top}$=0, which lead to $m_{h^0}=116$GeV. The cross section is very
sensitive to $f_{Ts}$ since the correction to the Higgs boson vertex
is one of the dominant corrections. When $f_{Ts}$ is smaller than 0.1,
the cross section is significantly suppressed. Since the box diagram
corrections to the scalar operators are numerically small, the $a_{\rm
eff}$ dependence is relatively small.

\begin{figure}
 \centerline{\epsfxsize = 0.7\textwidth \epsffile{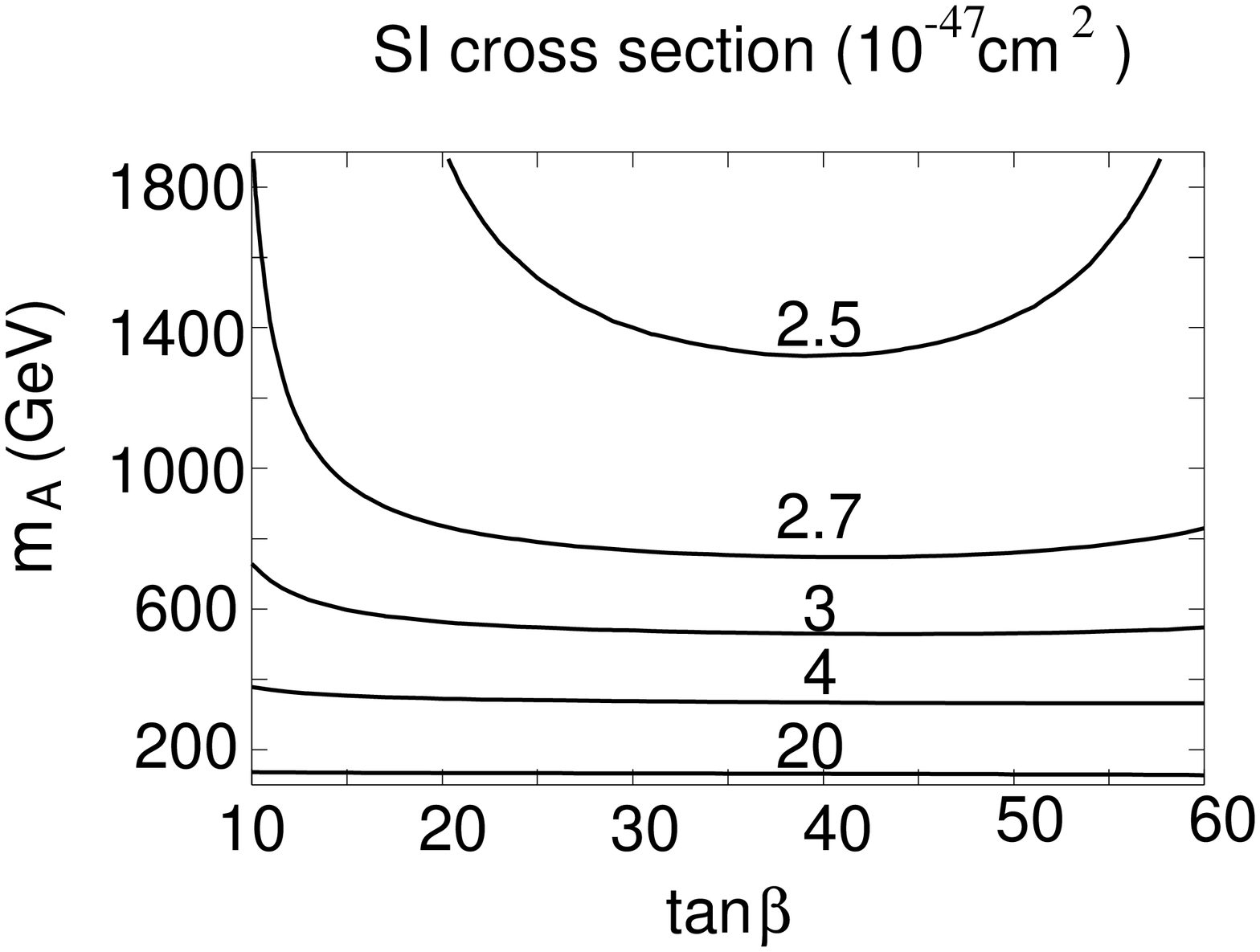} }
\vspace*{-5mm}
\caption{One-loop induced cross section for spin-independent
$\tilde{\chi}^0$--$p$ scattering is
presented as a function of $\tan\beta$ and $m_{A}$. The wino-like LSP
is assumed. Here, we take $m_{\tilde{\chi}^0}$=1600GeV, $m_{\rm stop}$=2000GeV,
$A_{\rm top}$=0, $a_{\rm eff}=1$, and the parameter in
Table.~\ref{table1}.  }
\label{figsi}
\end{figure}

In Fig.~\ref{figsi} the one-loop induced spin-independent cross section is
presented as a function of $\tan\beta$ and $m_{A}$. Here, we take
$m_{\tilde{\chi}^0_1}$=1600GeV, $m_{\rm stop}$=2000GeV, $A_{\rm top}$=0, $a_{\rm eff}=1$,
and the parameter in Table.~\ref{table1}. The $\tan\beta$ dependence
is moderate.  The one-loop induced  cross section rises as large as
$10^{-46}$cm$^{2}$ when the pseudoscalar Higgs boson is light.

In the case of the Higgsino-like LSP, the one-loop induced cross
section for the spin-independent scattering is smaller than that in
the wino-like case since the $SU(2)$ charge is smaller.
When the pseudoscalar Higgs boson and the LSP
are much heavier than the $W$ boson mass, $\delta f_N$ is given as
\begin{eqnarray}
\delta f_N
&=&
4.3\times 10^{-11}
-(5.1+ 0.1 a_{\rm eff})\times 10^{-11}
+6.0\times 10^{-11}
\nonumber\\
&=&5.1\times 10^{-11}, ~~~(a_{\rm eff}=1).
\end{eqnarray}
Here, we take $m_{h^0}=115$GeV, again. The order of the terms in the
first line is the same as Eq.~(\ref{order}). This value for $\delta f_N$
corresponds to $\sigma_{\rm SI}\simeq 1.1\times 10^{-48}{\rm cm}^2$
when the tree-level contribution is negligible.

Now we discuss the gauge-loop correction in the total cross section in
the general MSSM parameter space.  In above paragraphs we showed 
the cross section induced by the one-loop diagrams alone,  assuming the
tree-level cross section is negligible. When the tree-level amplitude
dominates, the one-loop correction to the cross section is
approximately expressed as $\sim 2
\sqrt{\sigma_{{\rm 1-loop}}/\sigma_{\rm tree}}$, where $\sigma_{\rm
tree}$ and $\sigma_{{\rm 1-loop}}$ are the tree-level and one-loop
induced cross sections, respectively. When $\sigma_{{\rm
1-loop}}/\sigma_{\rm tree}\sim 1/100$, the correction to the total
cross section is about 20\%.

\begin{figure}
\centerline{\epsfxsize = 0.7\textwidth \epsffile{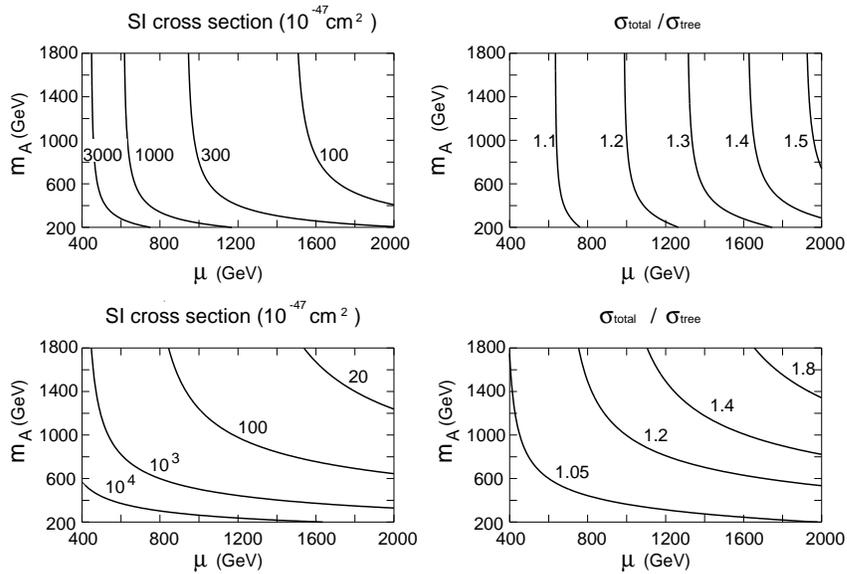}}
\caption{Total cross section for spin-independent
$\tilde{\chi}^0$--$p$ scattering, $\sigma_{\rm total}$, (left) for the
wino-like LSP in the $(\mu, m_A)$ plane. The cross section is
depicted in the unit of $10^{-47}$cm$^2$. Here, $M_2=200$~GeV,
$M_1/M_2=11t_W^2$, $m_{\tilde{t}}=2$~TeV and $\tan\beta=4(40)$ for
upper (lower) plot. $\sigma_{\rm total}$ includes the gauge-loop
contributions while the squark exchange contribution at tree level is
ignored.  In the right, we give the ratio $\sigma_{\rm
total}/\sigma_{\rm tree}$ for the same parameters.  $\sigma_{\rm
tree}$ is the cross section at tree level. }
\label{lightwino}
\end{figure}

\begin{figure}
\centerline{\epsfxsize = 0.7\textwidth \epsffile{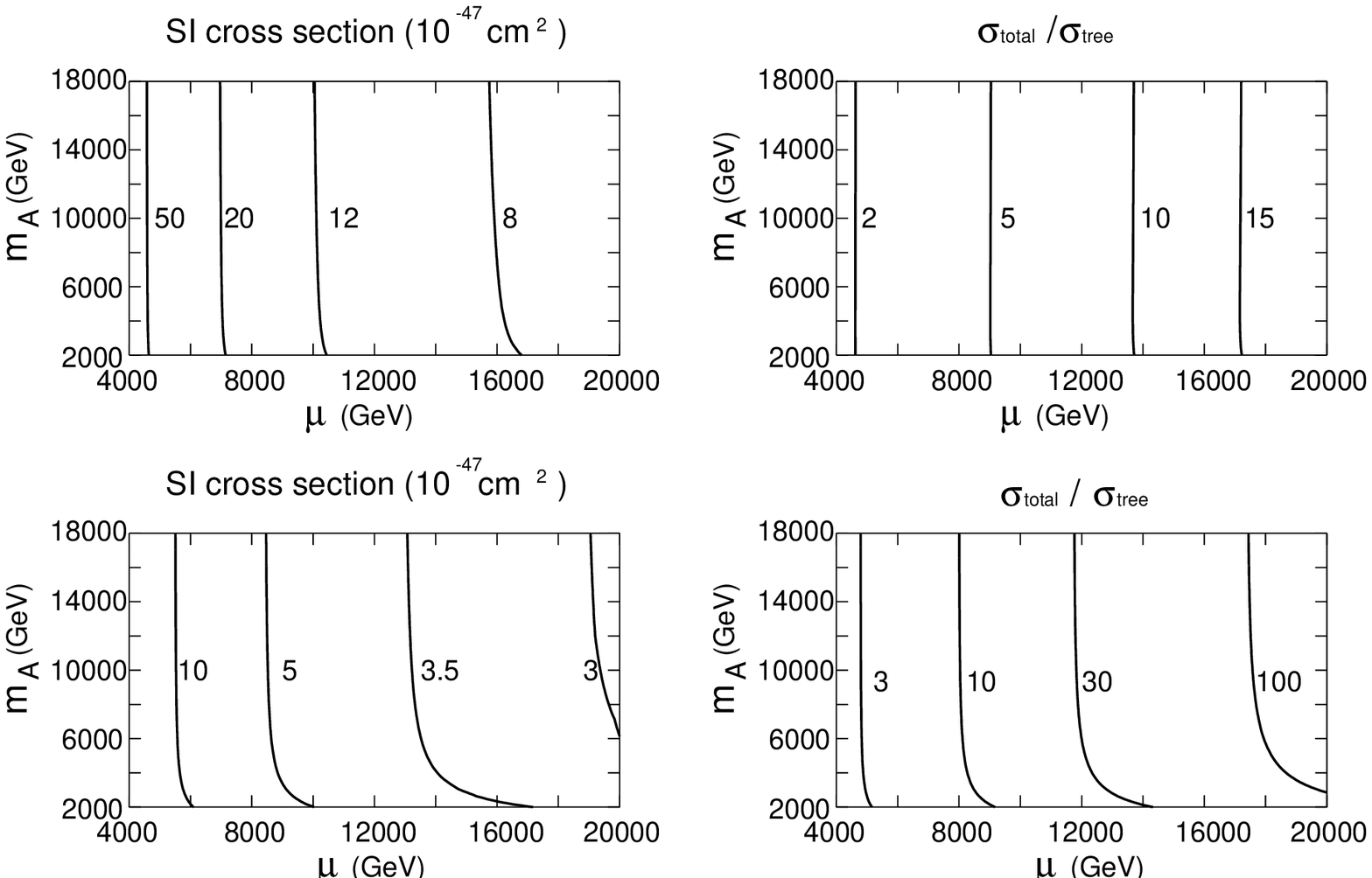}}
\caption{Same as Fig.\ref{lightwino}, but $M_2=2$~TeV, $m_{\tilde{t}}=20$~TeV.}
\label{heavywino}
\end{figure}

In Fig.~\ref{lightwino} we show the cross section involving the
tree-level and one-loop contributions, $\sigma_{\rm total}$, and
$\sigma_{\rm total}/\sigma_{\rm tree}$ for the spin-independent
$\tilde{\chi}^0$--$p$ scattering as as functions of $\mu$ and
$M_A$. Here, we assume the wino-like LSP and take
$m_{\tilde{\chi}^0}\sim200$GeV.  The upper two figures are for
$\tan\beta=4$ and the lower ones are for $\tan\beta=40$. See the
caption for the other input parameters.  It is found that the radiative
correction is about 50\% in the plots, where $\sigma_{\rm total}$ is
above $10^{-45}$~cm$^{2}$.  The radiative correction is relatively
significant in larger $\tan\beta$, since the coupling of the LSP with
the light Higgs boson at tree level is more suppressed in the case as
discussed in Section 3.1. In Fig.~\ref{heavywino}, we show the case where
$m_{\tilde{\chi}^0}\sim 2$~TeV. It is found that $\sigma_{\rm total}$
is less than $10^{-45}$~cm$^2$, however it is dominated by the
gauge-loop correction in the plots, since the LSP is very
close to the pure wino state.

\begin{figure}
\centerline{\epsfxsize = 0.7\textwidth \epsffile{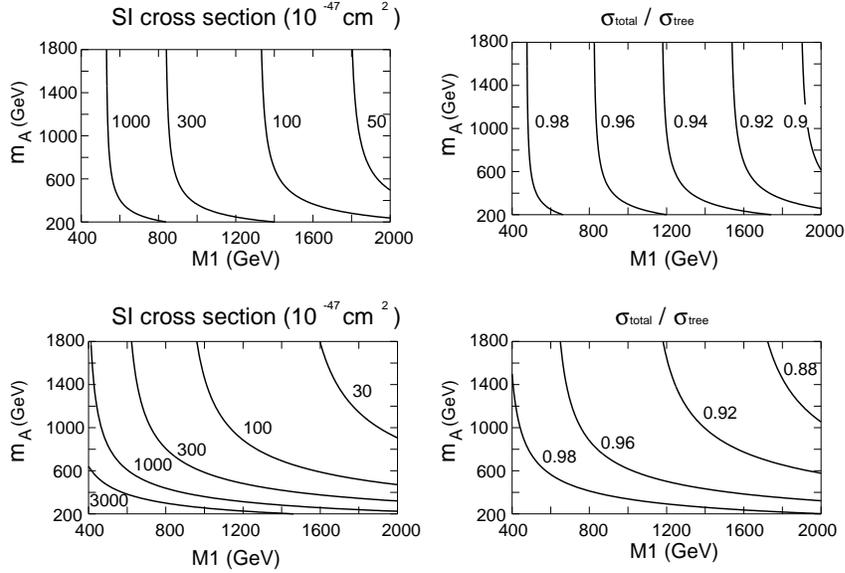}}
\caption{Total cross section for spin-independent 
$\tilde{\chi}^0$--$p$ scattering, $\sigma_{\rm total}$ (left) for the
Higgsino-like LSP in the $(\mu, m_A)$ plane. The cross section is 
depicted in the unit of $10^{-47}$cm$^2$. Here,
$\mu=200$~GeV. $M_1/M_2=5 t_W^2/3$, $m_{\tilde{t}}=2$~TeV and
$\tan\beta=4(40)$ for upper (lower) figure respectively. In the right,
we give the ratio $\sigma_{\rm total}/\sigma_{\rm tree}$.  }
\label{lighthiggsino}
\end{figure}

\begin{figure}
\centerline{\epsfxsize = 0.7\textwidth \epsffile{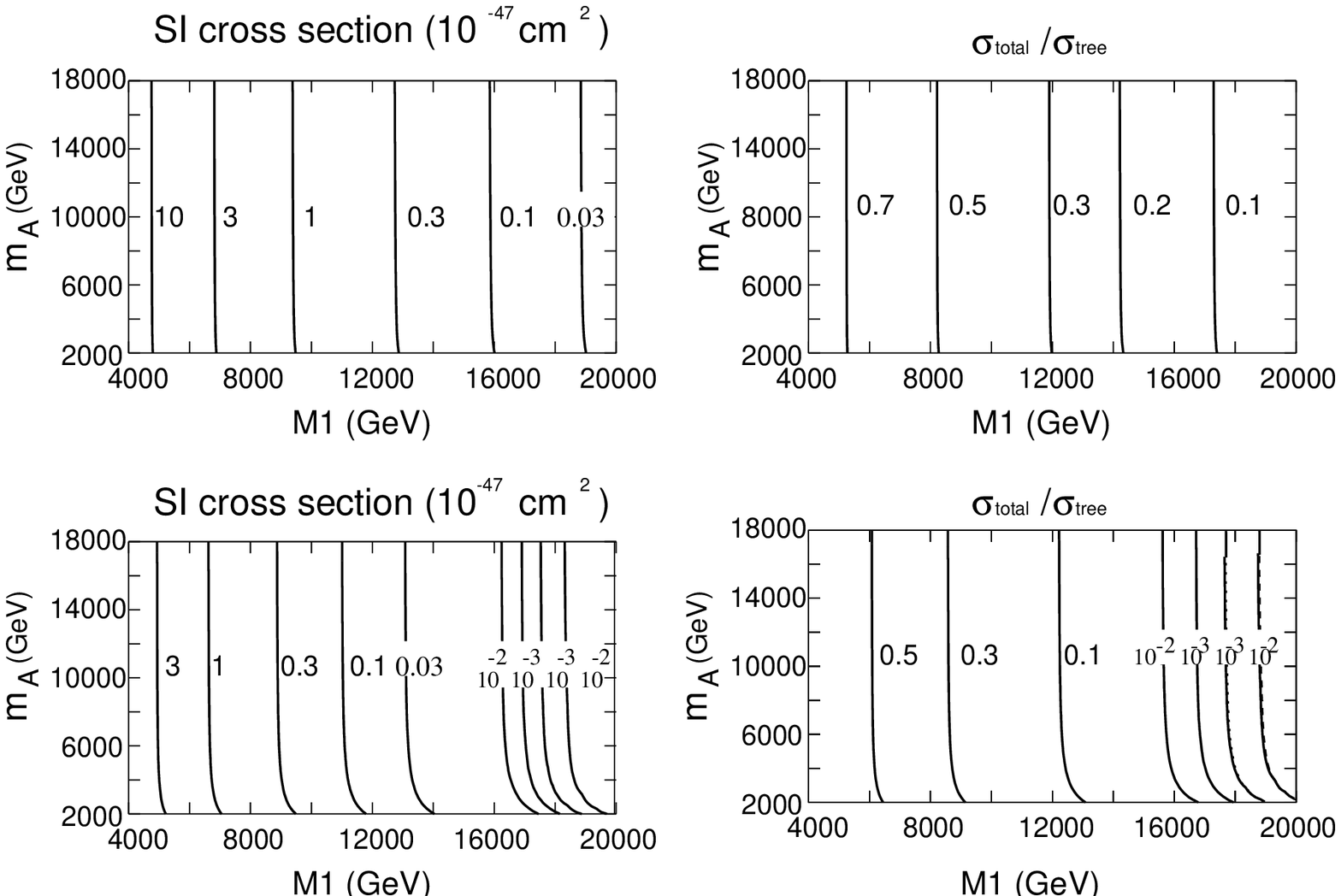}}
\caption{Same as Fig.~\ref{lighthiggsino}, 
but $\mu=2$~TeV and $m_{\tilde{t}}=20$~TeV.}
\label{heavyhiggsino}
\end{figure}

In Figs.~\ref{lighthiggsino} and \ref{heavyhiggsino}, we show the
effect of the radiative correction for the Higgsino-like LSP. The
radiative correction is less than $\sim 10$\% for
$m_{\tilde{\chi}^0}\sim 200$~GeV even when $M_A$ and $M_1$ are as
heavy as 2~TeV.  Compared with the wino-like case, the suppression of
the tree-level Higgs boson vertex by the gaugino-Higgsino mixing 
is more moderate while the radiative correction by the
gauge-loop diagrams is smaller.  The correction is negative (positive)
for $\mu>0(<0)$, relatively to the tree-level contribution. For
$m_{\tilde{\chi}^0_1}\sim 2$~TeV, the one-loop contribution cancels the
tree-level contribution to reduce the total cross section less than
$10^{-50}$~cm$^{2}$ in the figure.

\begin{figure}
 \centerline{\epsfxsize = 0.7\textwidth \epsffile{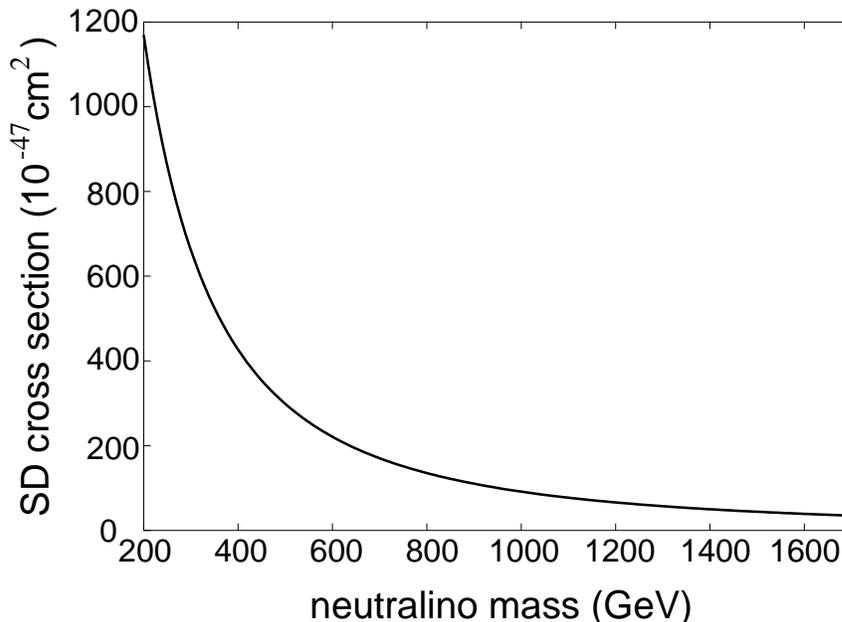} }
\vspace*{-5mm}
\caption{
One-loop induced cross section for the spin-dependent $\tilde{\chi}^0$--$p$
scattering in the case of the wino-like
neutralino DM. Here, we assume that the tree-level contribution is negligible.}
\label{figsd}
\end{figure}

Finally, we discuss the spin-dependent cross section. As discussed in the
previous section, the one-loop contribution to the process is
suppressed by $m_W/m_{\tilde{\chi}^0}$ in the amplitude, contrary to
the spin-independent cross section. In Fig.~\ref{figsd} the one-loop
induced cross section for the spin-dependent $\tilde{\chi}^0$--$p$
scattering is given as a function of the LSP mass under the assumption
of the wino-like LSP. In this figure we assume again that the
tree-level contribution is negligible.  The asymptotic behavior for
the cross section is
\begin{eqnarray}
\sigma_{\rm SD}
&\simeq&
\frac{\pi\alpha_2^4m_N^2}{3 m_{\tilde{\chi}^0}^2 m_W^2}
\left(\sum_{q=u,d,s} d_q\right)
\nonumber\\
&\simeq&
1.2\times 10^{-43}{\rm cm}^2\times \left(\frac{m_{\tilde{\chi}^0}}{100{\rm GeV}}\right)^{-2}.
\end{eqnarray} 
Thus, from Eq.~(\ref{sigmasdtreew}), it is found that the one-loop
correction becomes significant for $m_{\tilde{\chi}^0}\simeq 100$ GeV
when $\mu\gsim 1$~TeV. For $|\mu|\gg M_2$, the one-loop contribution is
constructive to the tree-level one for the spin-dependent
$\tilde{\chi}^0$--$p$ scattering while it is deconstructive for
the $\tilde{\chi}^0$--$n$ scattering.

For the Higgsino-like LSP, the one-loop induced spin-dependent cross
section is one order of magnitude smaller than that for the wino-like
LSP.  It behaves as
\begin{eqnarray}
\sigma_{\rm SD}
&\simeq&
1.5\times 10^{-44}{\rm cm}^2\times \left(\frac{m_{\tilde{\chi}^0}}{100{\rm GeV}}\right)^{-2}
\end{eqnarray} 
in a large $m_{\tilde{\chi}^0}$ limit. The tree-level contribution
does not suffer from significant suppression by the Higgsino-gaugino
mixing as in Eq.~(\ref{sigmasdtreeh}), and the one-loop correction is
negligible as far as the gaugino masses are smaller than 10~TeV. 
For $M_1$, $M_2 \gg|\mu|$, the
one-loop contribution is deconstructive (constructive) to the tree-level
one for the spin-dependent $\tilde{\chi}^0$--$p$
($\tilde{\chi}^0$--$n$) scattering.

\section{Conclusion and discussion}

In this paper, we studied one-loop correction to the
neutralino-nucleon scattering processes for the wino- and
Higgsino-like LSPs. The $\tilde{\chi}^0$--$N$ scattering is relevant to
the detection rate of the neutralino dark matter. The scattering occurs
dominantly through the exchange of the Higgs or $Z$ boson at tree
level, which is suppressed by the gaugino-Higgsino mixing. This is
especially the case when the gauginos or Higgsino mass is large.

On the other hand, the scattering cross section receives the one-loop
contribution from the processes involving the weak boson exchange.
These processes are not suppressed neither by the small mixing angle
nor by heavy SUSY mass scale, therefore it could be significant part
of the one-loop corrected cross section in the limits of the wino- and
Higgsino-like LSPs. The spin-independent cross section for the
wino-like (Higgsino-like) LSP receives the sizable one-loop
correction, when the Higgsino (gaugino) mass is heavier than about
1TeV and the spin-independent cross section is smaller than about
$10^{-45}$ cm$^2$ ($10^{-46}$ cm$^2$). 

There has been impressive progress on experimental techniques for the
dark matter searches. The on-going experiments are now sensitive
enough to exclude some of the MSSM parameters which may not be accessible
otherwise. Many proposals aim to push the sensitivities further. The
better determination on the $\tilde{\chi}^0$--$N$ cross section is
also needed to understand the local dark matter density and its
velocity distribution. Although we calculated only a part of the
correction of the scattering cross section, further investigation on
the loop effect might be needed in future.

\section*{Acknowledgment}
This work is supported in part by the Grant-in-Aid for Science
Research, Ministry of Education, Science and Culture, Japan
(No.15540255, No.13135207 and No.14046225 for JH and No.14540260 and
No.14046210 for MMN).

\section*{Appendix: Higgsino-like LSP}

In this Appendix, we present the radiative corrections to the effective
interactions for the LSP scattering with nucleon  when the LSP is
Higgsino-like.  The Higgsino-like LSP accompanies the chargino and the
second-lightest neutralino, whose masses are degenerate with that of
the LSP. Furthermore, the Higgsino-like LSP has an interaction with the $Z$
boson.

The one-loop corrections to the  LSP couplings with the Higgs bosons
are given as
\begin{eqnarray}
\delta c_{h\tilde{\chi}\tilde{\chi}}[\tilde{\chi}^-]
&=&
\frac14 \frac{\alpha_2}{4\pi}
\sin({\alpha-\beta})
\left[F_{\rm H}^{(0)}(x_W)
+
\delta_C
F_{\rm H}^{(1)}(x_W)
\right]
\nonumber\\
&&-\frac14 \frac{\alpha_2}{4\pi c_W^2}
\sin({\alpha-\beta})
\left[F_{\rm H}^{(0)}(x_Z)
+
\delta_N
F_{\rm H}^{(1)}(x_Z)
\right],
\\
\delta c_{H\tilde{\chi}\tilde{\chi}}[\tilde{\chi}^-]
&=&
-\frac14\frac{\alpha_2}{4\pi}
\cos({\alpha-\beta})
\left[F_{\rm H}^{(0)}(x_W)
+
\delta_C
F_{\rm H}^{(1)}(x_W)
\right]
\nonumber\\
&&+\frac14\frac{\alpha_2}{4\pi c_W^2}
\cos({\alpha-\beta})
\left[F_{\rm H}^{(0)}(x_Z)
+
\delta_N
F_{\rm H}^{(1)}(x_Z)
\right].
\end{eqnarray}
The contributions from the box diagrams to the coefficients in the
effective Lagrangian (\ref{boxeff}) are approximated as
\begin{eqnarray}
\delta d_q[\boxx]&=&\frac{1}{4}\frac{\alpha_2^2}{m_W^2}
\left[F_{\rm AV}^{(0)}(x_W)+\delta_C F_{\rm AV}^{(1)}(x_W)
\right]
\nonumber\\
&&
-\frac12 (L_q^2+R_q^2) \frac{\alpha_2^2}{c_W^2 m_W^2} 
\left[
F_{\rm AV}^{(0)}(x_Z)
+
\delta_N F_{\rm AV}^{(1)}(x_Z)
\right],
\end{eqnarray}
and 
\begin{eqnarray}
\delta f_q[\boxx] &=&
(L_q R_q) \frac{\alpha_2^2}{c_W m_W^3}
\left[F_{\rm S1}^{(0)}(x_Z)
+
\delta_N F_{\rm S1}^{(1)}(x_Z)
\right],
\label{zi}
\\
\delta f'_q[\boxx] &=&
\frac{1}{4}\frac{\alpha_2^2}{m_W^3}
\left[F_{\rm S2}^{(0)}(x_W)
+
\delta_C F_{\rm S2}^{(1)}(x_W)
\right]
\nonumber\\
&&-\frac12(L_q^2+R_q^2) \frac{\alpha_2^2}{c_W m_W^3}
\left[F_{\rm S2}^{(0)}(x_Z)
+
\delta_N F_{\rm S2}^{(1)}(x_Z)
\right],
\label{zii}
\\
\delta g_q^{\rm (I)} &=&
\frac{1}{4}
\frac{\alpha_2^2}{m_W^3}
\left[F_{\rm TI}^{(0)}(x_W)
+
\delta_C F_{\rm TI}^{(1)}(x_W)
\right]
\nonumber\\
&&-\frac12(L_q^2+R_q^2) \frac{\alpha_2^2}{c_W m_W^3}
\left[F_{\rm TI}^{(0)}(x_Z)
+
\delta_N F_{\rm TI}^{(1)}(x_Z)
\right],~~~~({\rm I}=1,2).
\label{ziii}
\end{eqnarray}
Here, $L_q=T_3^q-Q_q s_W^2$ and $R_q=-Q_q s_W^2$ for a  quark $q$.
The $Z$ boson contributions appear in the above equations. 
In the equations, $x_Z=m_Z^2/m^2_{\tilde{\chi}^0}$,
$\bar{b}_Z=\sqrt{1-x_Z/4}$, and $\delta_N\equiv (m_{\tilde{\chi}_2^0}
-m_{\tilde{\chi}^0})/m_{\tilde{\chi}^0}$ with $m_{\tilde{\chi}_2^0}$
the heavier neutralino mass. All mass functions are the same as in
Eqs.~(\ref{massfunc}) except for $F_{\rm S1}^{(0)}(x)$ and $F_{\rm
S1}^{(1)}(x)$,
\begin{eqnarray}
F_{\rm S1}^{(0)}(x)
&=&
\frac{1}{4\bar{b}_Z}
(4-x(2-x))\tan^{-1}(\frac{2\bar{b}_Z}{\sqrt{x}})
+
\frac14 \sqrt{x}(2-x \log(x)),
\\
F_{\rm S1}^{(1)}(x)
&=&
\frac{3}{4\bar{b}_Z^3}(2-x)
\tan^{-1}(\frac{2\bar{b}_Z}{\sqrt{x}})
\nonumber\\
&&-
\frac{1}{4 \bar{b}_Z^2}\frac{1}{\sqrt{x}}
(4-4x-2(4-x)\log(2\delta_N)+(4-x)\log(x))
.
\end{eqnarray}

\end{document}